\documentclass[twoside]{elsart}
\usepackage[dvips]{graphicx}
\usepackage{amssymb}

\newcommand{\dg}{\ensuremath{^{\circ}}}

\newcommand{\mpn}{A.MeV}
\newcommand{\qpi}{\ensuremath{4\pi}}

\newcommand{\gu}{\nuc{155}{Gd} + \nuc{nat}{U}}
\newcommand{\xe}{\nuc{129}{Xe} + \nuc{nat}{Sn}}

\newcommand{\moy}[1]{\mbox{$\langle #1\rangle$}}
\newcommand{\mutot}{\mbox{$N_{C}$}}
\newcommand{\nf}{\mbox{$N_{f}$}}

\newcommand{\mlcp}{\mbox{$N_{LCP}$}}
\begin{document}

\begin{frontmatter}

\title
{Multifragmentation of a very heavy nuclear system (II): bulk properties
 and spinodal decomposition    
\thanksref{Ganil}} 
\thanks[Ganil]{Experiment performed at Ganil}
\author[ipno]{J.D.~Frankland\thanksref{pres-ganil}},
\author[ipno]{B.~Borderie},
\author[cata]{M.~Colonna},
\author[ipno]{M.F.~Rivet},
\author[ipno]{Ch.O.~Bacri},
\author[ganil]{Ph.~Chomaz},
\author[lpc]{D.~Durand},
\author[ipno]{A.~Guarnera},
\author[buch]{M.~P\^arlog},
\author[ipno]{M.~Squalli},
\author[buch]{G.~T\u{a}b\u{a}caru}, 
\author[ganil]{G.~Auger},
\author[lpc]{N.~Bellaize},
\author[lpc]{F.~Bocage},
\author[lpc]{R.~Bougault},
\author[lpc]{R.~Brou},
\author[cea]{P.~Buchet},
\author[ganil]{A.~Chbihi},
\author[lpc]{J.~Colin},
\author[lpc]{D.~Cussol},
\author[cea]{R.~Dayras},
\author[ipnl]{A.~Demeyer},
\author[cea]{D.~Dor\'e},
\author[ipno,cnam]{E.~Galichet},
\author[lpc]{E.~Genouin-Duhamel},
\author[ipnl]{E.~Gerlic},
\author[ipnl]{D.~Guinet},
\author[ipnl]{P.~Lautesse},
\author[ganil]{J.L.~Laville},
\author[lpc]{J.F.~Lecolley},
\author[cea]{R.~Legrain},
\author[lpc]{N.~Le Neindre},
\author[lpc]{O.~Lopez},
\author[lpc]{M.~Louvel},
\author[ipnl]{A.M.~Maskay},
\author[cea]{L.~Nalpas},
\author[lpc]{A.D.~Nguyen},
\author[ipno]{E.~Plagnol},
\author[napol]{E.~Rosato},
\author[ganil]{F.~Saint-Laurent\thanksref{cada}},
\author[ganil]{S.~Salou},
\author[lpc]{J.C.~Steckmeyer},
\author[lpc]{B.~Tamain},
\author[ipno]{L.~Tassan-Got}, 
\author[ganil]{O.~Tirel},
\author[lpc]{E.~Vient}, 
\author[cea]{C.~Volant},
\author[ganil]{J.P.~Wieleczko}

\collaboration{INDRA collaboration}

\address[ipno]{Institut de Physique Nucl\'eaire, IN2P3-CNRS, F-91406 
Orsay Cedex, France.}
\address[cata]{Laboratorio Nazionale del Sud, Viale Andrea Doria, 
  I-95129 Catania, Italy}
\address[ganil]{GANIL, CEA et IN2P3-CNRS, B.P.~5027, F-14076 Caen Cedex,
  France.}
\address[lpc]{LPC, IN2P3-CNRS, ISMRA et Universit\'e, F-14050 Caen 
  Cedex, France.}
\address[buch]{National Institute for Physics and Nuclear Engineering,
RO-76900 Bucharest-M\u{a}gurele, Romania}
\address[cea]{DAPNIA/SPhN, CEA/Saclay, F-91191 Gif sur Yvette Cedex,
France.}
\address[ipnl]{Institut de Physique Nucl\'eaire, IN2P3-CNRS et 
  Universit\'e, F-69622 Villeurbanne Cedex, France.}
\address[cnam]{Conservatoire National des Arts et M\'etiers, F-75141 Paris
Cedex 03, France.}
\address[napol]{Dipartimento di Scienze Fisiche e Sezione INFN, 
  Universit\`a di Napoli ``Federico II'', I80126 Napoli, Italy.}
\thanks[pres-ganil]{Corresponding author.\\
Permanent address: GANIL, B.P.~5027, F-14076 Caen Cedex, France.\\
E-mail: frankland@ganil.fr. Tel.: 33~231~454628. Fax: 33~231~454665}
\thanks[cada]{present address: DRFC/STEP, CEA/Cadarache, F-13018
Saint-Paul-lez-Durance Cedex, France.}

\begin{abstract}

The properties of fragments and light charged particles emitted in
multifragmentation of single sources formed in central 36 \mpn{} Gd+U
collisions are reviewed. Most of the products are isotropically distributed
in the reaction c.m. Fragment kinetic energies reveal the onset of radial
collective energy. A bulk effect is experimentally evidenced from the
similarity of the charge distribution with that from the lighter 32 \mpn{}
Xe+Sn system. Spinodal decomposition of finite nuclear matter exhibits the
same property in simulated central collisions for the two systems, and
appears therefore as a possible mechanism at the origin of
multifragmentation in this incident energy domain.

\end{abstract}



\end{frontmatter}

\section{Introduction}

The decay of highly excited and possibly compressed nuclear systems
through multifragmentation (complete disassembly into several lighter 
nuclei in a 
short time scale) is, at present time, a subject of great interest in
nucleus-nucleus collisions. While this process has been observed for 
many years, its experimental knowledge was strongly improved only 
recently with the advent of powerful \qpi\ devices; the major 
experimental problem comes from the difficulty to unambiguously select
and well define the relevant system  or subsystem. Part I of the present 
work fully illustrates this fact when selecting fused
systems formed in collisions between very heavy nuclei~\cite{gadoue-i}.

Our goal in studying multifragmentation of very heavy systems, which
can be considered as well defined pieces of nuclear matter, was to
seek out bulk properties which could be compared to models in which 
bulk or volume instabilities are present.

Many theories have been developed to explain multifragmentation (see 
for example ref.~\cite{MO93} for a general review of models). In 
particular one can arrive at the concept of multifragmentation by
considering a  phase transition of excited nuclear matter. This 
first phase transition is of liquid-gas type due to the specific 
form of the nucleon-nucleon interaction, which is characterized by 
attraction at long and intermediate range and repulsion at short range. 
It is possible that, during a collision between two nuclei, a wide zone 
of the nuclear matter phase diagram may be explored, including 
the liquid-gas phase coexistence region and even more 
precisely the spinodal region (domain of negative incompressibility and 
of mechanical instability of uniform matter) where multifragmentation 
can occur through the growth of density fluctuations~\cite{BE83}.

Among the existing models of multifragmentation some are related to
statistical approaches based either on multi-body phase space
calculations~\cite{GRO90,BO95,LO89,RA97} or on fast sequential binary 
decays~\cite{FR90}, whereas others try to describe the 
dynamic evolution of systems resulting from collisions between two 
nuclei. In the present case, semi-classical simulations based on the nuclear
Boltzmann equation, which describe the time evolution of the one-body density,
are appropriate during the early phase of the collisions but become
inadequate when instabilities occur. A quantum-mechanical description 
including N-body correlations is not yet feasible, thus 
dynamic scenarios taking into account the dynamics of the phase
transition are simulated, with different approximations, via molecular 
dynamics~\cite{PE89,FE90,ONO93,ONO96,SU99} or stochastic mean field 
approaches~\cite{AY88,RAN90,CHO91}. 

It is this last approach which has been used for a comparison with our
experimental observables. In simulations, spinodal decomposition of hot 
and dilute finite nuclear systems is mimicked through a powerful tool,
the Brownian One-Body (BOB) dynamics~\cite{CH94,GU96,GU97}.
The BOB dynamics reintroduces approximately N-body correlation effects (i.e.
fluctuations) by means of a brownian force in the mean field.
The magnitude of the force is adjusted to produce the same r.m.s. 
fluctuation amplitude as the full Boltzmann-Langevin theory for the
most unstable modes in nuclear matter prepared at the corresponding
density and temperature~\cite{CHO96}.

The paper is organized as follows : the experimental results concerning
selected multifragmenting fused events from 36 \mpn{} \gu\ collisions are 
presented in section~\ref{SSE}; the dynamical properties of fragments 
and their connection with an eventual expansion energy are discussed 
in section~\ref{SIMON}; in section~\ref{XeGd} a bulk effect is brought 
to light by comparison of experimental fragment charge and 
multiplicity distributions 
with results from a lighter system. Section~\ref{BOB} is devoted to a
presentation of the dynamical simulations and their confrontation with 
the experimental data. A summary is given in section~\ref{conc}.

\section{Characteristics of single-source events}\label{SSE}

Single-source events, for which very heavy fused systems are formed from 
the majority of nucleons of a heavy projectile and target, have been 
isolated among very well measured 36 \mpn{} $^{155}Gd$+$^{238}U$ 
collisions (see Part I). Multifragmentation of these sources 
comprising more than 350 nucleons leads to fragments with charges up 
to 60. For such heavy reaction 
products recombination effects in solid state detectors (silicon
detectors and CsI(Tl) scintillators) are important, and thus a great 
effort was recently made to obtain more accurate energy calibrations 
(globally within 6\% over the whole detection angular range) 
and Z identification procedures (see Part I)~\cite{TA99,PA00}.
 
Before presenting in detail the characteristics of the isolated fused 
systems, let us first recall very briefly what we learned in Part I 
(Section 4). Single-source events are selected by imposing the condition 
$Z_{tot}$$\geq$120 ($\approx$0.77$Z_{sys}$) for charge completeness,
and through a global shape variable, namely the polar angle 
$\theta_{flow}$, calculated from the emission properties of fragments 
(the detection of at least three fragments with Z$\geq$5 was required).
From the forward-backward symmetry of reaction product Z-distributions 
observed in these selected events, we can infer that the single-source 
has nearly the centre of mass velocity. In this section we will first 
present and discuss observables related to fragments. Then properties 
of light charged particles (LCP) associated to these events, which do 
not enter the chosen selection, will be presented and used to derive
an estimate of the characteristics (size and excitation energy) 
of the source. 

\begin{figure}[htbp!]
\begin{center}
\includegraphics[height=.75\textwidth]{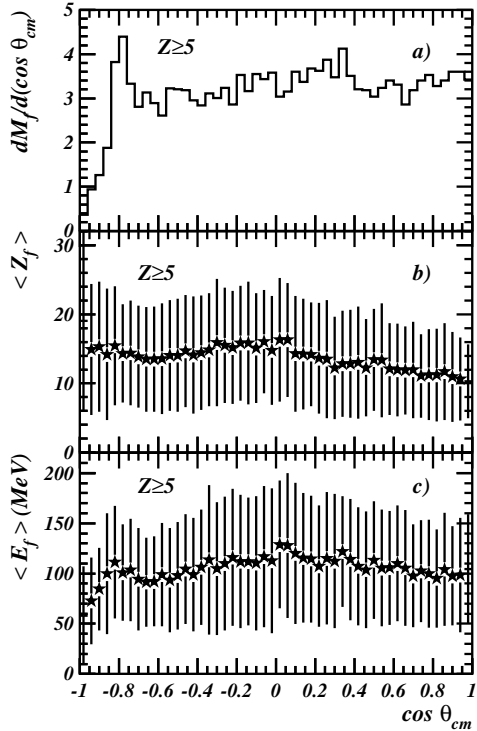}
\caption{\textit{a) c.m. angular distribution of fragments emitted
by the single source formed in central 36 \mpn{} \gu{}. b) Average
fragment atomic number versus their c.m. emission angle; c)Average c.m. 
fragment kinetic energy versus their c.m. emission angle. Bars in b) and c) 
refer to measured standard deviations.
\label{da_f}}}
\end{center}
\end{figure}

\subsection{Fragment emission properties}\label{SSE-frag}

Figure~\ref{da_f}a) shows, in the reaction centre of mass frame, 
the angular 
distribution of emitted fragments. It is isotropic, except at large 
angles ($\theta_{cm} > 135^o$ ) where detection and identification 
thresholds come into play. On figures~\ref{da_f}b) and c)  are displayed 
the evolution of fragment properties (average Z and average kinetic 
energy) with respect to their emission angle in the centre of mass;
these average values are constant within $\pm$15\%; The slight increase 
around 90$^o$ probably comes from the selection of events 
($\theta_{flow}\geq70^o$ - see Part I)).
\begin{figure}[htbp!]
\begin{center}
\includegraphics[width=.75\textwidth]{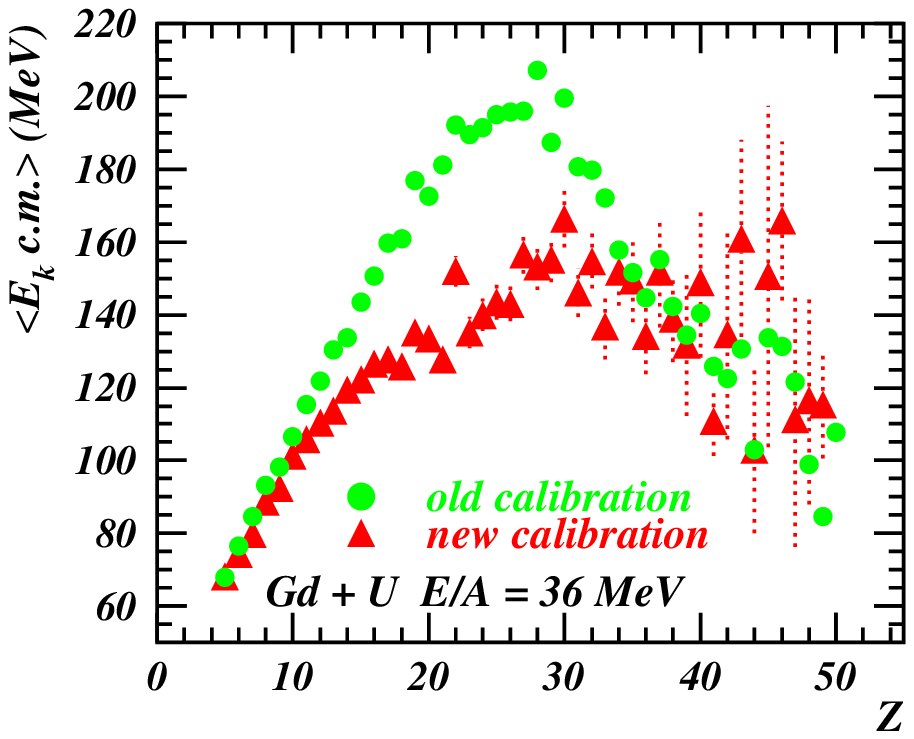}
\caption{\textit{Average c.m. fragment kinetic energy versus their atomic
number for the single source formed in central 36 \mpn{} \gu{} collisions. 
New CsI calibrations are used (triangles). Previous data as
published in~\cite{RI98} are also shown (circles). Statistical error bars
are shown only for new calibrations. \label{ek_on} }}
\end{center}
\end{figure}

Figure~\ref{ek_on} shows the evolution of the average kinetic energy of
fragments versus their atomic number.
The energy first increases with Z, up to 150 MeV at Z$\sim$30. Then it
levels off, or slightly decreases.
In the same figure are also reported previously published data
~\cite{RI98} which did not take into account recent improvements in the
energy calibration of the backward CsI(Tl) scintillators of 
INDRA~\cite{PA00}; differences are essentially observed for intermediate
charges, Z=15-30. This correlation between kinetic energy and Z of 
fragments is of particular interest and will be discussed later on in 
the paper: it can indeed permit the extraction of qualitative 
information about fragment emission (sequential or simultaneous), 
and quantitative information related to radial expansion energy 
(section~\ref{SIMON}). Consequently
it puts strong constraints on theoretical models (section~\ref{BOB}).
\begin{figure}[htbp!]
\begin{center} 
\includegraphics[width=.75\textwidth]{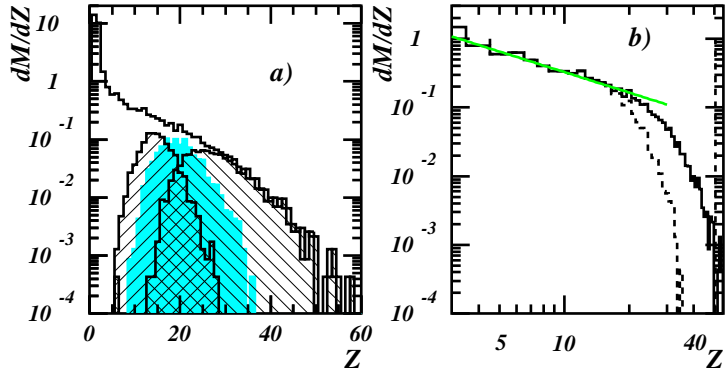}
\caption{\textit{a):Differential charged product multiplicity
distributions for the single source formed in central 36 \mpn{} \gu{}
collisions. The hatched (grey) histograms show the Z distributions of
the 3 largest fragments. b): Same with abscissa in logarithmic scale,
for all fragments (solid histogram), excluding the largest (dashed 
histogram). The line shows a fit with a power law in $Z^{-0.99}$.
\label{dmdz_f} }}
\end{center}
\end{figure}

The differential multiplicity as a function of the atomic number of the
detected charged reaction products is presented in figure~\ref{dmdz_f}a.
It extends up to Z$\sim$55, which corresponds to a third of the total 
charge of the system. The contributions to the 
distribution of the three heaviest fragments of each partition are also 
displayed in the figure, showing that only the heaviest one populates 
the region Z=35-60. In figure~\ref{dmdz_f}b the distribution for 
fragments is plotted in a double logarithmic scale;
a power law dependence in $Z^{-\tau}$ of the differential multiplicity 
becomes evident in the range Z=5-20, with $\tau$ close to 1.0.
Finite size effects break this law for higher 
charges, and the heaviest fragment is completely excluded from this 
dependence.
Note that such a behaviour is predicted by simulating the explosion of 
hot drops of classical fluid in the spinodal region, but with a somewhat
higher exponent~\cite{PR95}.
\begin{figure}[htbp!]
\begin{center}
\includegraphics[width=.75\textwidth]{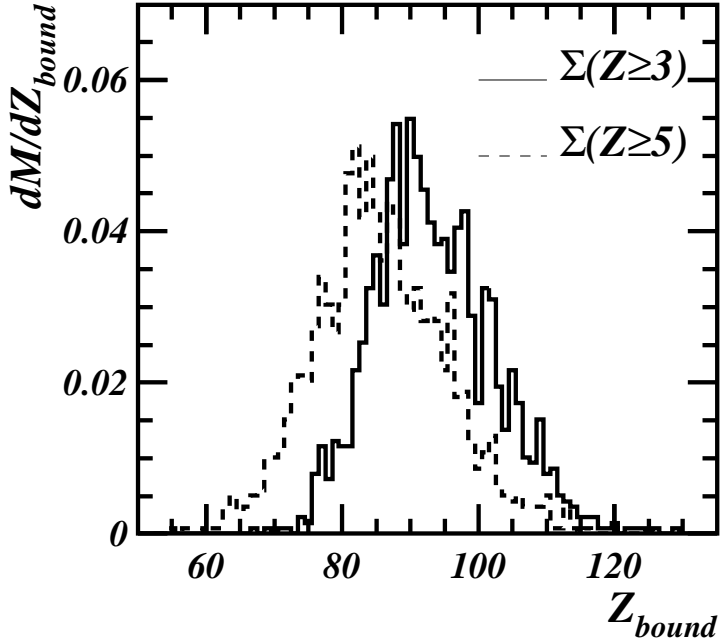}
\caption{\textit{Distribution of the charge bound in fragments for the
single-sources produced in central 36 \mpn{} \gu{} collisions.\label{zbound}}}
\end{center}
\end{figure}
The distribution of 
the total charge emitted in fragments (Z$_{bound}$) is displayed in
Figure~\ref{zbound}; on average, more than half of the total charge of 
the system is bound in fragments. Finally Table~\ref{su_chrs_1}
summarizes some average measured quantities related to fragments 
emitted by the single source (the different multiplicity distributions 
associated to single-source events have been presented in Fig. 10 of 
the accompanying paper).

\begin{table}[htbp!] 
\begin{center}
\caption{\small Average measured values of total, light charged particle and 
fragment multiplicities, of the total charge bound in fragments 
$Z_{bound}$, of the fragment atomic number, and of the atomic number of the 
3 largest fragments of each event. Except in columns 1 and 2, only fragments
with $Z\geq 5$ are considered. \label{su_chrs_1}}
\begin{tabular}{cccccccc} 
\hline 
\moy{\mutot} & \moy{\mlcp} & \moy{\nf} & \moy{Z_{bound}} & 
\moy{Z} & \moy{Z_{max1}} & 
\moy{Z_{max2}} & \moy{Z_{max3}} \\ 
\hline 
33.2 & 24.5 & 6.3 & 86.5 & 14.2 & 26.9 & 18.8 & 14.0\\ 
\hline 
\end{tabular} 
\end{center} \end{table} 

\subsection{Coincident LCP and estimates of average size and excitation energy
  of the source}\label{SSE-lcp}

\begin{figure}[htbp!]
\begin{center}
\includegraphics[width=.75\textwidth]{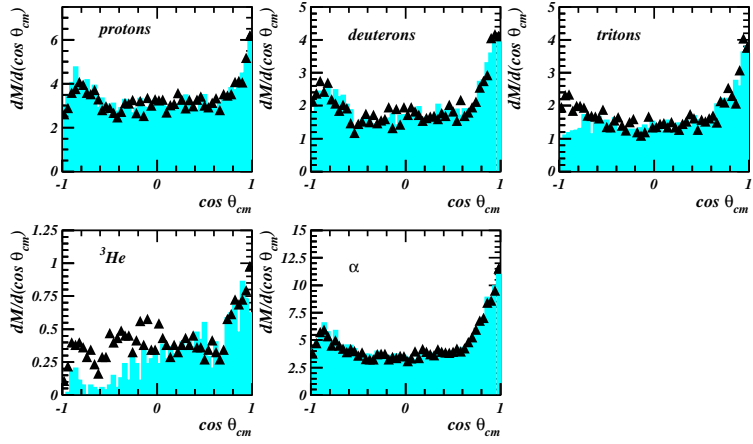}
\caption{\textit{c.m. angular distribution of light charged particles
emitted in coincidence with the single source formed in central
36 \mpn{} \gu{}. Histograms are raw data, while triangles
correspond to corrected values (see text).\label{da_lcp} }}
\end{center}\end{figure}

LCP are emitted at different stages during the collisions. First
``direct'' pre-equilibrium emission occurs which partially keeps a 
memory of the entrance channel mass asymmetry~\cite{FU94}. Then 
particles are emitted before and during the formation of fragments 
from a source which may be in thermal
equilibrium. Finally in the late stages statistical evaporation from
hot fragments takes place. Experimentally we measure the sum of all these
contributions and, without having recourse to correlation
functions~\cite{MA98}, only deviations from isotropy in the centre of mass
 can be used to give complementary information about the source.
\begin{figure}[htbp!]
\begin{center}
\includegraphics[width=.75\textwidth]{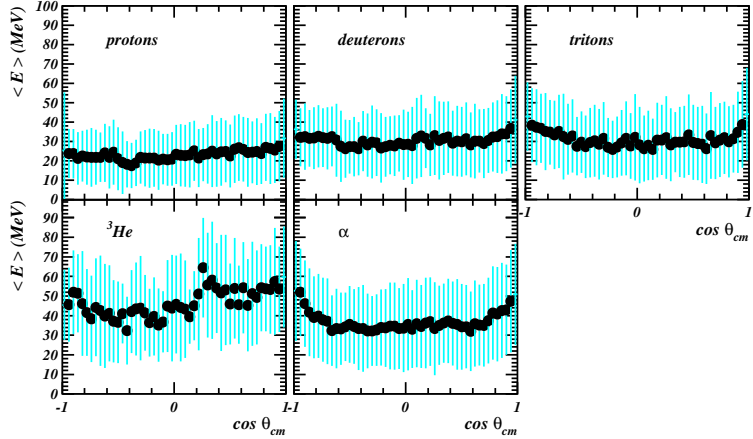}
\caption{\textit{c.m. average energy vs the c.m. angle of light charged
particles emitted in coincidence with the single source formed in 
central 36 \mpn{} \gu{} (corrected values). \label{ek_lcp} }}
\end{center}
\end{figure}

The angular distributions of LCP and their associated average kinetic energy
as a function of the centre of mass angle are displayed in figures~\ref{da_lcp}
and~\ref{ek_lcp}.
Deuterons, tritons and $^3He$  suffer from isotopic
identification thresholds; their contributions at low laboratory kinetic
energies were estimated from the measured H and He by linearly extrapolating
isotope ratios (d/H, t/H, $^3$He/He) measured as a function of the particle
laboratory kinetic energy and angle. Corrected values correspond to
triangles in figure~\ref{da_lcp}. The angular distributions for all LCP 
are very similar, being flat between 60\dg{} and 120\dg{}, and presenting 
forward and backward peaks, the latter being lower (even if detection 
thresholds would be accounted for). Angular momentum
effects would lead to angular distributions symmetric with respect to
90\dg{}, and unrealistically large values of spin would be required 
to explain the measured anisotropy for protons between 0\dg{} and 90\dg{} 
for instance. Other arguments against a very
high angular momentum of the emitter will be given in section~\ref{SIMON}.
Consequently we consider that two contributions are superimposed:
a forward-backward direct component (the forward-backward asymmetry
arising from the asymmetry of the incident nuclei~\cite{FU94}) and
an isotropic one which dominates in the angular range 60\dg{}-120\dg{}.
A maximum estimate of the isotropic contribution coming from the
single-source is obtained by doubling the multiplicities measured in this
angular range. Its characteristics are summarised
in Table~\ref{chrs-equilibre}. The relative abundances of the heavy hydrogen
isotopes are higher than those currently observed for evaporated particles 
(M$_t$/M$_p \sim$0.1); this indicates that particle emission becomes
isotropic rather early in the collision process, and more particularly well
before hot and well-separated fragments deexcite through evaporation.

\begin{table}[htbp!]
\begin{center}
\caption{\small Characteristics of ``equilibrated'' LCP emission:
Average multiplicities ($= 2 \times$ the average multiplicity of particles
emitted between  60\dg{} and 120\dg{}); average c.m. kinetic energies;
inverse slope of energy spectra $\tau $ (see text).
\label{chrs-equilibre}}
\begin{tabular}{cccccc}
\hline
\ & $p$ & $d$ & $t$ & \nuc{3}{He} & $\alpha$ \\
\hline
$\moy{M_{eq}}$ & 6.0 & 3.4 & 2.8 & 0.8 & 7.3\\
\hline
$\moy{E^{eq}_{c.m.}}$ (MeV)& 22.0 & 29.5 & 29.0 & 46.1 & 34.4\\
\hline
$\tau$ (MeV)& 12.82 & 15.54 & 14.68 & 19.83 & 15.77\\
\hline
\end{tabular}
\end{center} \end{table}

\begin{figure}[htbp!]
\begin{center}
\includegraphics[width=.75\textwidth]{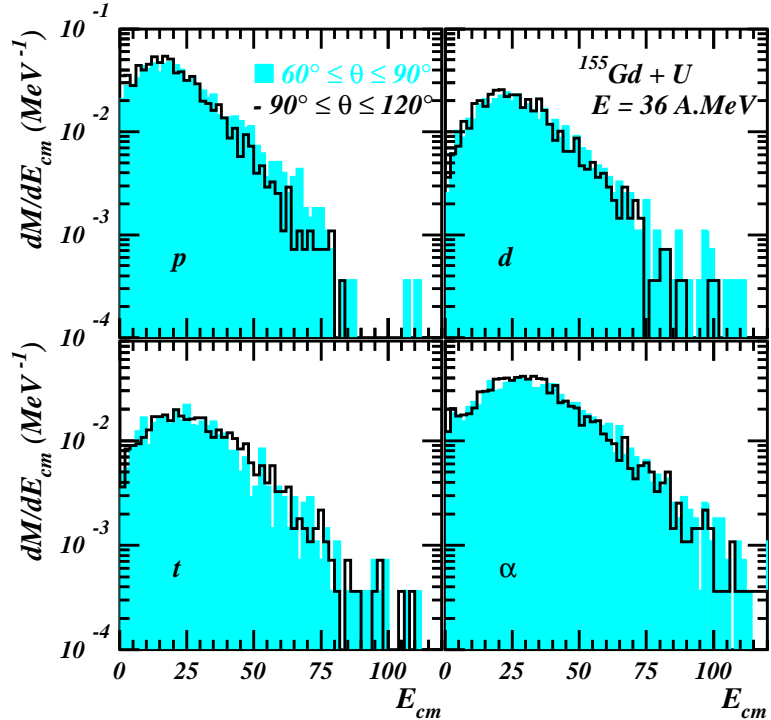}
\caption{\textit{c.m. light charged particle energy spectra for two 
angular ranges. \label{sp_lcp} }}
\end{center}
\end{figure}

Energy spectra of isotropic LCP emitted in the range 60\dg{}-90\dg{} and
90\dg{}-120\dg{} in the centre of mass are presented in
Figure~\ref{sp_lcp}. These spectra are, as expected, identical for the
two defined angular ranges. However they cannot be fitted by any
statistical formula (surface or volume emission). This fact
suggests again that these spectra result from
emissions at different stages of the reaction, preventing us from 
inferring any information relative to the source temperature.
However  we can characterize the isotropic emissions by average
quantities (multiplicity and kinetic energy)  and slope parameters  $\tau$
determined from the high-energy exponential fall-off of the spectra
(Table~\ref{chrs-equilibre}). The hierarchy of the average lcp energies
is a general feature of central collisions between very heavy ions:
d and t have equal average energies and slope parameters, much higher
than the corresponding proton quantities, and closer to that for $\alpha$'s; 
the $^3He$ average energy is in turn spectacularly higher than
the $\alpha$ energy~\cite{BOU99}. The same features are observed in the INDRA Xe+Sn data
between 32 and 50 \mpn{}, and in the FOPI Au+Au data between 150 and 250
\mpn{}~\cite{RE97}. A possible interpretation would be that d,t and $^3He$ 
mostly arise from the multifragmentation stage, while p and $\alpha$ are 
in addition abundantly evaporated by colder and colder fragments, thus 
lowering drastically their global average energy.

The characteristics of the direct (anisotropic) emission were 
estimated by subtracting the isotropic contribution determined above
from the total distributions. 
Average values for multiplicities and energies of direct particles are
reported in Table~\ref{preeq}; the indicated neutron multiplicity, was 
deduced by assuming that the anisotropic emission keeps the N/Z of the total
system. Neutron energy was derived from proton energy by subtracting a
Coulomb barrier.

\begin{table}[htbp!]
\begin{center}
\caption{\small Average multiplicities and kinetic energies
of preeequilibrium particles for single-source events, estimated following
the method described in text.\label{preeq}}
\begin{tabular}{ccccccc}
\hline
\ & $n$ & $p$ & $d$ & $t$ & \nuc{3}{He} & $\alpha$ \\
\hline
$<M_{pe}>$ & 3.16 & 0.60 & 0.64 & 0.68 & -  & 2.0\\
\hline
$<E^{pe}_{CM}>$ (MeV) & 23.0 & 37.7 & 39.0 & 38.6 & -  & 49.5\\
\hline
\end{tabular}
 \end{center}\end{table}

The detection efficiency of INDRA for LCP being excellent (90\% of 
\qpi{}), we relied on the LCP rather than on the measured fragments 
to estimate the single source characteristics:
upper limits for the charge, mass and excitation energy of the 
single source were derived by subtracting the quantities removed 
by direct particles (without any efficiency correction) 
(Table~\ref{preeq}) from the corresponding values for the total system.
The average mass and charge of the source are then $<A>=378$, $<Z>=150$,
which corresponds to 96\% of the total system. Its average
excitation energy  is 6.5 MeV per
nucleon to be compared with the 7.1 MeV available for the
composite system in the centre of mass.

\section{Properties of the single-source: sequential decay or
 simultaneous break-up? Expansion energy}\label{SIMON}

Properties of multifragment systems are frequently derived from
comparison with statistical models. The aim is to determine whether 
thermodynamical equilibrium has been reached at some stage of the collision, 
through the ensemble of partitions experimentally observed. These models 
are purely static, in the
sense that they describe the system at a freeze-out instant, defined as the
time where the nuclear interaction between the fragments vanishes.
Fragment kinetic energy only comes from their mutual Coulomb repulsion and 
thermal motion. Any deviation of the measured kinetic energy
from the predicted values is then attributed to some extra collective
energy, such as expansion energy, or rotational energy. 
For instance a self-similar expansion, decoupled from thermal motion, is
generally added in order to reproduce the measured kinetic energies.
The values of collective energy so derived are however obviously dependent 
on the Coulomb energy, or in other words on the volume assumed at
freeze-out~\cite{MI96,ADN98}. The Statistical Model for 
Multifragmentation (SMM~\cite{BO95}), and the 
Microcanonical Metropolis Monte-Carlo model (MMMC~\cite{GRO90}) are 
widely used, for their correct statistical weighting of partitions. 
The SIMON event generator~\cite{ADN98} is chosen here for two reasons.
Firstly its highly simplified
algorithm for generating partitions reasonably accounts for the
measured charge and multiplicity distributions. Secondly, and above
  all SIMON has the advantage 
over the models cited above to permit a rigorous treatment of
space-time correlations between all fragments and emitted particles.

In SIMON, the inputs are: the mass, charge and available energy of the system;
the number and spatial configuration of the primary fragments and 
eventually their minimum mass;
the radial expansion energy at the rms radius. The hot fragments are then
propagated while deexciting (through the transition state formalism in the
present case).
Statistical deexcitation of a hot ``compound'' nucleus can also be
followed, by setting to 1 the number of primary fragments. 
An additional physical feature was recently included in this code, 
namely a variation of the level density with the excitation energy: the 
level density is indeed expected to vanish at high excitation~\cite{DE85}.
This is equivalent to excluding from the primary partitions levels with 
too short life-times. The formalism adopted here is that proposed
in~\cite{KO87}, where the level density at energy $\varepsilon$ is expressed
as the Fermi gas level density modified by a modulation factor.

\begin{equation}
\rho^{eff}(\varepsilon) = \rho^{FG}(\varepsilon) \times e^{-\varepsilon/T_{lim}}
\end{equation}

This corresponds to having an effective intrinsic nuclear temperature:

\begin{equation}
 T^{-1}_{eff} = T^{-1} + T^{-1}_{lim}
\end{equation}

Therefore the available thermal energy of the system, after removing 
Coulomb and collective parts, is shared between kinetic and intrinsic 
excitation energy of the $M_f$ fragments, following the equation:

\begin{equation}
E = 3M_fT/2 + \sum_1^{M_f} a T_{eff}^2
\end{equation}

$a$ being the level density parameter at zero temperature (taken here 
as $a$=A/10).

\begin{figure}[htbp!]
\begin{center}
\includegraphics[width=.75\textwidth]{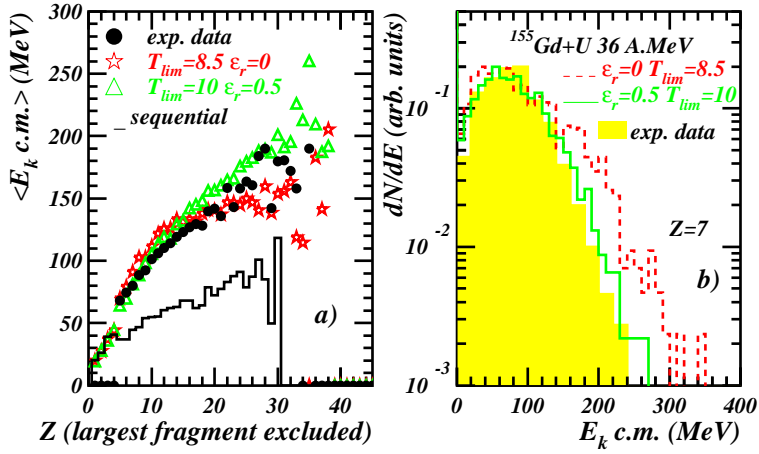}
\caption{\textit{a): average c.m. fragment kinetic energies versus their
charges (the heaviest fragment is excluded). 
Experimental data (points) are compared to different SIMON
simulations. b): kinetic energy spectra for Z=7; the filled histogram
represent experimental data, and the solid and dashed lines result from
SIMON simulations. $\varepsilon_r$ is in \mpn{} and T$_{lim}$ in MeV.
\label{simon}}}
\end{center}
\end{figure}

The initial system was chosen as that defined in section~\ref{SSE-lcp},
after removing direct particles.
A first simulation was performed to test the assumption of sequential
deexcitation of the hot source as the origin of fragment emission.
Indeed a modified sequential scenario based on the Expanding Emitting Source
model was shown to account for some features of the system considered
here~\cite{FR99}. With the SIMON simulation, the calculated charge
distribution does not extend to large enough values, and the
fragment kinetic energies are largely underestimated, as appears in
Fig~\ref{simon}a.
In the following simulations, multifragmentation of the initial system
in 6 primary fragments, with a minimum mass of 20, was assumed.
Indeed, when only looking at kinematic variables, the hypotheses of
light fragments and particles being produced in the explosion process
itself, or evaporated by very hot and spatially close fragments, are
expected to give nearly identical results. No angular momentum was
considered as it was found that, for this very heavy system, an initial
value of 500~$\hbar$ (corresponding to b = 4 fm) would increase by less
than 5 MeV the final fragment kinetic energy~\cite{FR98}. In the two
simulations shown in Fig~\ref{simon}, it was verified that the calculated
multiplicity and charge distributions reasonably account for the
experimental ones, without attempting a perfect fit. The
temperature $T_{lim}$ and the collective energy were thus determined by
adjusting the calculated fragment kinetic energies on the experimental
data. $T_{lim}$ may be expected to lie between 5 and 12 MeV:
5 MeV is the temperature obtained from most data based on excited state
population ratios, and also from a correlation study between particles
and fragments for single sources formed in 50 \mpn{} Xe+Sn
collisions~\cite{MA98}. 12 MeV is the maximum nuclear temperature predicted
for nuclei when considered as liquid drops in equilibrium with their
vapor~\cite{LE85}. Without collective energy, the derived value of 
$T_{lim}$ is equal to 8.5 MeV; in this case the experimental variation 
of the fragment kinetic energy (including the largest one) with their 
charge is almost perfectly reproduced, independently of how the fragments 
are initially sitting in space.
As soon as expansion energy comes into play, selected initial spatial
configurations must be chosen, with the heaviest fragment close to
the centre of gravity, in order to follow the experimental saturation of 
the kinetic energy at high Z (see Fig~\ref{ek_on}). The rising part of 
the curve E(Z) is however independent of the spatial configuration, and 
a way to avoid a configuration choice is to consider all fragments but 
the largest, as in Fig~\ref{simon}a. An equally good reproduction of
experimental data is obtained with $T_{lim}$=10 MeV and 
$\varepsilon_r$ = 0.5 \mpn{}; the initial excitation energy of the 
fragments in this case is 4.9 \mpn{}. The maximum collective energy
compatible with the data is 
1 \mpn{} if $T_{lim}$ is increased to 12 MeV.
Fragment spectra were then scrutinized, in order to choose between the
hypotheses with 0 or 0.5 A.MeV expansion energy. 
In Fig~\ref{simon}b is shown the experimental
spectrum for Z=7, and the calculated spectra with and without expansion
energy. Clearly the hypothesis with expansion energy 
is the best, the option $\varepsilon_r$ = 0 
leading to a too broad spectrum. Note that the discrimination between the 
two assumptions can only be made for light fragments (Z$<$10), as for 
charges Z=12-20 the two calculated spectra are almost superimposable.

In conclusion of this section, the comparison of measured fragment 
kinetic energies with a simultaneous break-up scenario 
reveals the need for a radial collective motion at the freeze-out 
whose energy is $\varepsilon_r$ = 0.5 $\pm$ 0.5 \mpn{}.
This indicates that expansion energy begins to appear for central 
collisions between very heavy ions around 30 \mpn{}, as also found for a
similar system~\cite{MI96}. The origin of this expansion (compression,
thermal pressure?) will be discussed in section~\ref{BOB}, with the help of
dynamical simulations.

\section{Experimental evidence for a bulk effect: the fragment  charge 
distribution is independent of the charge of the total system.}\label{XeGd}

\begin{figure}[htbp!]
\begin{center}
\includegraphics[width=.75\textwidth]{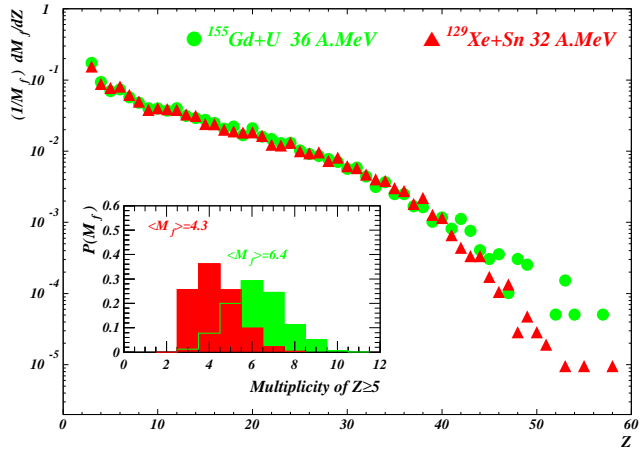}
\caption{\textit{ Experimental differential charge multiplicity distributions,
normalised to each event's M$_f$, and fragment (Z$\geq$5) multiplicity (M$_f$)
distributions (insert) for the 32 \mpn{} Xe+Sn (black 
histogram and triangles) and 36 \mpn{} Gd+U (grey histogram and circles) 
systems. Average experimental multiplicity values are given. 
From~\cite{RI98}. \label{gdxeexp}}}
\end{center}
\end{figure}

\xe{} reactions between 25 and 50 \mpn{} were also studied with the INDRA
array~\cite{RI98,PLA99,MA97,SA97,NLN99}. From 32 \mpn{} up, compact single 
source events could be isolated with the same 
flow angle selection as explained in the accompanying paper.
At 32 \mpn{}~\cite{RI98}, the available excitation energy per nucleon 
for the total
system is the same as for the \gu{} system studied in the previous 
sections. An experimental effect, observed for the first time, stands out
for  single-source multifragmentation events from these two systems 
(Fig~\ref{gdxeexp}): the charge distributions (normalised by the 
average multiplicity of Z$\geq$3) superimpose over 3 orders of 
magnitude while the average fragment multiplicities are in 
the ratio 1.49, i.e. almost exactly the ratio of the total charges of 
the systems 156/104. Such a peculiar behaviour may reveal that
bulk effects play a major role in the multifragmentation of these systems. 
Note that the same Z distribution is also observed in ref.~\cite{MI96} 
for the 35 A.MeV Au+Au system, which has a total charge close to \gu{}.

As shown in section~\ref{SIMON}, the kinematical properties of 
single-source events are compatible with the assumption that, 
at some instant of the collision a freeze-out configuration is reached, 
i.e. the system can be described as an 
ensemble of (hot) charged products and neutrons in thermodynamical 
equilibrium, free from mutual nuclear interaction. This implies that 
the total system  at this instant occupies a much larger volume than would
a nucleus with the
same mass, or in other words that it has reached a low (though 
inhomogeneous) density state. Statistical models like MMMC, SMM do 
predict that two systems with the same thermal energy (temperature) will, 
on average, break up into a number of fragments proportional to their
mass, with the same fragment charge distributions~\cite{LEF97}. In 
these models the charge (mass) of the largest fragment essentially 
reflects the excitation energy; indeed
the measured average charges of these fragments only differ by 7\%  
for the two systems studied in this section, 32 \mpn{} \xe\ and 36
\mpn{} \gu{} (Fig~\ref{BOB2}).

The dilute systems considered in the statistical models have, in the
experiments, been produced through violent nuclear collisions,
and have lived quite a long history before reaching this stage. How such
a low density state is attained should bring information on the role of 
fundamental properties of nuclear matter such as incompressibility, 
viscosity \ldots For similar systems of colliding heavy ions, some transport
models for instance predict that in central collisions below 50 \mpn{},
an irreversible evolution towards low density is obtained after a 
compression phase if the incompressibility parameter of nuclear matter 
is close to 220 MeV; conversely for larger K$_{\infty}$ the system is less
strongly compressed and returns to normal density~\cite{MO92}. 
In other calculations however, multifragmentation in central Pb+Au 
collisions appeared to be driven by neutron/proton asymmetry rather 
than by the softness or stiffness of
the equation of state~\cite{JOU95}.

Instabilities of different kinds (volume, surface \ldots )
which are dynamically generated in the course of the collision may 
be suspected of causing the break-up of the system~\cite{MO93}. For 
the heavy systems considered here ($\sim$ 250 and 400 nucleons), bulk
instabilities may particularly be expected.

\section{Comparison with a stochastic mean-field model}\label{BOB}

The family of dynamical simulations of nuclear collisions based on the
nuclear Boltzmann equation, the 
Landau-Vlasov (LV), Boltzmann-Uehling-Uhlenbeck (BUU) or
Boltzmann-Nordheim-Vlasov (BNV) codes, were very successful in 
accounting for a
variety of experimental findings~\cite{CA90,DLM92,BA92,CO95,BO97,BA98}.
However, as they follow the time evolution of the one-body density, 
neglecting higher than binary correlations,
they ignore fluctuations about the mean trajectory of the system,
which becomes a severe drawback if the system happens to explore
regions of instability such as the spinodal zone, for instance. 
Indeed, in the presence of instabilities, local density fluctuations are
propagated and amplified by the mean-field leading to a multi-fragment 
breakup of the system. This is the so-called spinodal decomposition 
scenario for multifragmentation. The spinodal decomposition described in
ref.~\cite{GU96}, predicts a ``primitive'' break-up into fragments 
with a favoured size, connected to the wave lengths of the
most unstable modes in nuclear matter. These wave lengths are roughly 
constant as long as the systems are large enough. 
Therefore break-up into equal-sized fragments is expected, 
with a multiplicity growing with the mass of the system. This very simple
picture is in reality blurred by several effects: the beating of
different modes, the coalescence of fragments while the nuclear
force still acts and the finite size of the system will tend to end-up
with a broad, exponentially falling distribution of fragment charges.
Therefore it first appeared easier to determine whether spinodal 
decomposition would predict, for two different systems, the effect 
described in section~\ref{XeGd} rather than to look directly for a 
trace of spinodal properties in a single system. 

Let us now  come in more detail to the predictions of stochastic 
mean-field simulations of nucleus-nucleus collisions, based on 
the Boltzmann-Langevin (BL) equation, which allows for the treatment of 
unstable systems. Since the application of the BL theory to
3D nuclear collisions is still too computer demanding for calculating
realistic scenarios and for quantitative results, methods allowing to
simulate approximately the dynamical path followed by nuclear systems
crossing the spinodal region were developed. They consist of a ``Boltzmann
evolution'' starting from an inhomogeneous initial system~\cite{GU96,CO94} 
(Stochastic Initialisation Method: SIM)  or  complemented by a brownian
force~\cite{CH94,GU97} (Brownian One-Body Dynamics: BOB). In either case 
the amplitude of the initial density fluctuations or the magnitude of the 
stochastic force is chosen to reproduce the dynamics of the most unstable 
modes for infinite nuclear matter in the spinodal region.

A first comparison between experimental data and calculation was
published in reference~\cite{RI98}; it was found that the
multiplicity and charge distributions for both systems - and thus the
identical fragment charge distributions- were correctly reproduced, but 
that the calculated fragment
kinetic energies were too low. In this previous paper however, the
poorest of the two methods for simulating the BL equation proposed
was used (SIM): it consisted in initializing the correct (classical) 
density fluctuations when the system enters the spinodal region,
but continuing afterwards a standard one-body calculation, which 
unfortunately leads to  some  damping of fluctuations even when the 
mean field is unstable, because of the lack of a source term. 
As a result, the fragment formation 
time was incorrectly increased, leading to smaller final kinetic
energies because of the decrease with time of the expansion velocity of 
the unstable source. The second, and better, simulation method  proposed
(BOB) will be used for the comparisons presented in the
following~\cite{FR98}: the density fluctuations are now continuously 
created through the addition to the mean field of a stochastic force, 
and therefore maintained all along the evolution of the unstable system. 
A second major improvement has been implemented in the simulations: 
quantal fluctuations connected with collisional memory effects are 
now taken into account, as calculated in~\cite{AY94}, with the 
determinant result of doubling the overall amplitude of fluctuations 
for the most unstable modes in our case.
The dynamics of head-on 32 \mpn{} \xe\ and 36\mpn{} \gu{} collisions was
repeated with the improved simulations.  The method used to select our 
experimental data restricts to events which are highly relaxed in form 
(see Part I), representing only a part of all the multifragmenting 
single-source events~\cite{LEF99,BOU00}. Therefore only collisions at zero
impact parameter were simulated, shape effects are neglected. It should 
be noticed that although the noise amplitude is computed to match the 
fluctuation of bulk instabilities, it may trigger any type of instabilities.

In a coherent treatment of the BL theory, fluctuations should be
implemented from the beginning of the reaction, though their role 
only becomes crucial when instabilities appear. Indeed, they then 
lead to bifurcations and an ensemble of identically-prepared systems 
will explore a large number of different dynamical trajectories, which 
produce different multi-fragment partitions. Thus the stochastic
mean field calculation has to be performed as many times as the number
of events which one wishes to generate, which is very computer 
time-consuming for large statistics. Moreover the BOB dynamics is only 
applicable to locally-equilibrated systems and so cannot give a correct 
description of the very earliest, far off-equilibrium stages of the 
collision. Compromises therefore have to be made. As the BNV
calculation of the reaction without fluctuations gives a unique evolution, 
it was performed only once and the results used as initial conditions for 
the BOB dynamics. In order to correctly simulate the growth rate of 
fragments in the unstable system, the amplitude of density fluctuations 
must be correct at the time when the spinodal region is reached.
Then the BOB calculation has to begin before the onset of instability,
because the brownian force cannot set up density fluctuations of the 
correct amplitude instantaneously, due to the test-particles' inertia 
and finite relaxation times.
We chose as starting point the moment of maximum compression of the 
system (t=40 fm/c, see Table 4), when the local equilibrium is 
established, and we verified that the characteristics of the system at 
the onset of instability in the BOB calculation were identical to those 
found from the `true' one-body evolution of the reaction calculated with BNV.

\begin{figure}[htbp!]
\begin{center}
\includegraphics[width=.75\textwidth]{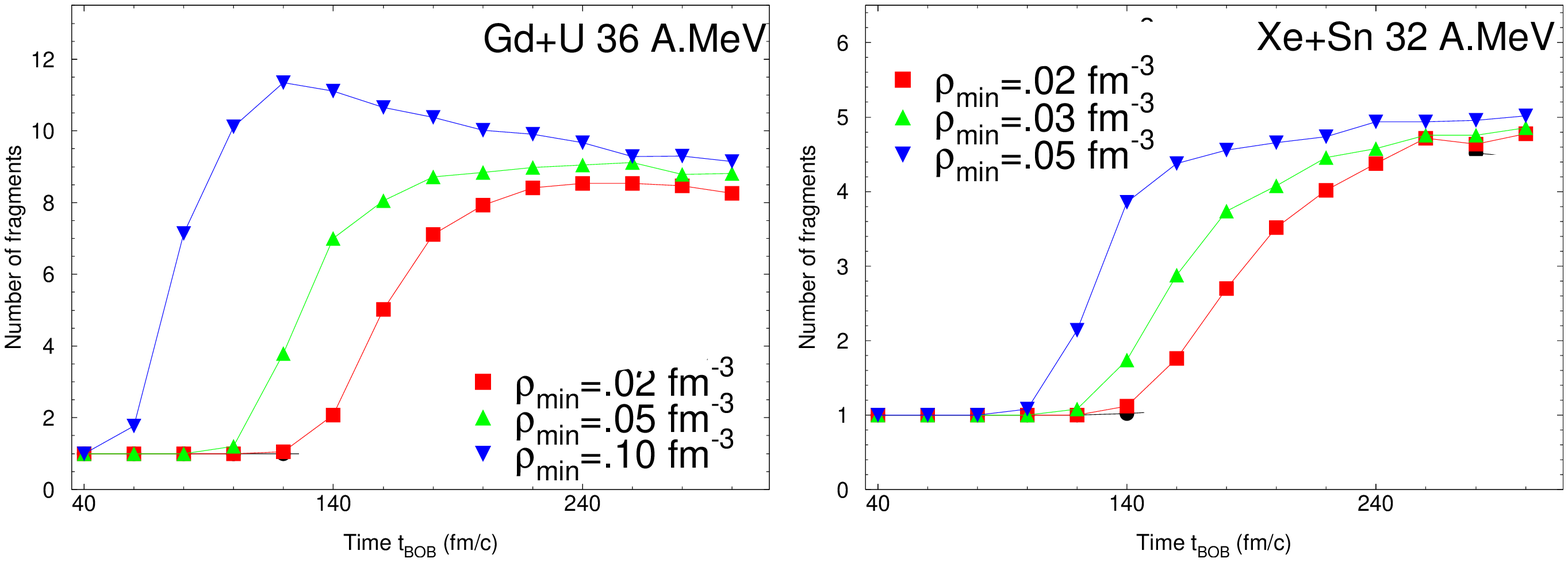}
\caption{\textit{Evolution with time of the average fragment (Z$\geq 5$)
multiplicity calculated for Gd+U and Xe+Sn collisions, for different
density cut-off~\cite{FR98}. \label{mf_BOB}}}
\end{center}\end{figure}

The ingredients of the BNV/BOB simulations are as follows. The
self-consistent mean field potential~\cite{ZA73} chosen gives a 
soft equation of state (K$_{\infty}$= 200 MeV) and the
finite range of the nuclear interaction is taken into account using a
convolution with a gaussian function with a width of 0.9 fm~\cite{GU96}. 
The addition of a term proportional to $\Delta \rho$  in the mean-field 
potential allows to well-reproduce the surface energy of ground-state 
nuclei~\cite{gat}. This is essential in order to correctly describe the 
expansion dynamics of the composite source.  In the collision term a constant
$\sigma_{nn}$ value of 41 mb, without in-medium, energy, isospin or 
angle dependence is used~\cite{BE78}. It should be noticed that, using a stiff 
parameterization of the equation of state (K$_{\infty}$= 380 MeV), 
the dynamical evolution of the composite source follows a very different 
path: density oscillations remain small and the system does not enter 
the spinodal region. 
Characteristics of the Gd+U and Xe+Sn  systems at two
times are shown in Table~\ref{tab_bnv}.
\begin{table}[htbp!]
\begin{center}
\caption{\small Left: t=40 fm/c, time of maximum compression; ($^a$)
Local temperature at the center of the non-equilibrated system.
Right: t=100 fm/c: thermalised systems inside the spinodal zone.
$v_{max}/c$ is the radial velocity at the surface. \label{tab_bnv}}
\begin{tabular}{rccccccccc}
\hline
\ & \multicolumn{4}{c}{Maximum Compression } &
\multicolumn{5}{c}{Spinodal zone} \\
\ & $A$ & $Z$ & $\rho/\rho_0$ & $T$ & $A$ & $Z$ &
$\rho/\rho_0$ & $T$ & $v_{max}/c$ \\
\ &  &  &  & (MeV) &  &  &  & (MeV) & \\
\hline
\xe\ & 247 & 103 & 1.25 & 8.3($^a$) & 238 & 100 & 0.41 & 4.0 &.09\\
\gu\ & 389 & 154 & 1.27 & 8.3($^a$) & 360 & 142 & 0.41 & 4.0 &.10\\
\hline
\end{tabular}
\end{center}
\end{table}
A maximum density  25-30\% higher than normal density is reached after
40 fm/c, corresponding to a moderate compression. Then during their
expansion phase the two systems enter the spinodal region at around
80 fm/c and attain slightly later thermal equilibrium at low density,
with a temperature of 4 MeV. The radial velocity at the
surface ($\sim$0.1c) reveals the gentle expansion of the systems and
the density fluctuations have time to develop leading to the formation
of fragments. For Gd+U, the charge and mass of the source at 100 fm/c
are close to the values determined for the experimental single-source
emitting isotropically particles and fragments (end of Sect.\ref{SSE}).
An algorithm for reconstructing fragments is applied at intervals of
20 fm/c, based on a minimum density cut-off. The calculation is stopped
when the fragment multiplicity becomes constant and independent of
reasonable variations of the value of the cut-off density, provided
it is larger than 0.01 fm$^{-3}$, as shown in Fig.~\ref{mf_BOB}. 
The chosen density cut-off was 0.02 fm$^{-3}$ for \gu{}; because of the
weaker Coulomb energy and radial flow for \xe{} collisions (see
table~\ref{tab_BOB}), which lead to  a different event topology at freeze-out, 
a higher density cut-off of 0.05 fm$^{-3}$  was used for this system.
The different characteristics of the fragments so defined can be
calculated: mass, charge, linear and angular momentum, kinetic energy
and excitation energy. A temperature is deduced from the relation
T=$\sqrt{E^*/a}$ using the density dependence of $a$ in the framework 
of the Fermi gas model.
The calculated kinetic energy of fragments does not contain the thermal
part. This is due to the fact that the amplitude of the BOB term is
tuned in order to reproduce the dynamics of the most unstable modes.
Hence the fluctuations that are introduced act mostly on the
configuration space, leading to fragmentation, but in principle
they do not automatically provide a sufficiently good description of
fluctuations in momentum space. The approximate solution to
this problem used here is to add to each component of the fragment
velocity a random part, distributed according to a gaussian of
width $\sigma^2 = T/M$, where $M$ is the mass and $T$  the average 
internal temperature of the fragment considered. This results in an 
increase of the average kinetic energy of fragments of $3/2~T$. 
The excitation energy of fragments is then reduced  
by the same quantity, in order to conserve the total energy. 
Attention is also paid to momentum conservation, event by event.
  
\begin{table}[htbp!]
\begin{center}
\caption{\small Characteristics of the systems at the end of the BOB
simulations, when starting the de-excitation step. $A_{tot}$ and
$Z_{tot}$ are the average mass and charge shared between $<N_f>$
fragments ($Z\geq5$) of average atomic number $<Z_f>$. \label{tab_BOB}}
\begin{tabular}{rccccccc}
\hline
\ & $A_{tot}$ & $Z_{tot}$ & $<N_f>$ & $<Z_f>$ & $<\varepsilon^*>$ &
$<\varepsilon_{rad}>$ & $E_{coul}$ \\
\ &  &  &  &  & (A.MeV) & (A.MeV) & (MeV) \\
\hline
\xe\ & 194.0 & 76.1 & 5.1 & 13.4 & 3.2 & 0.81 & 175.2\\
\gu\ & 320.0 & 120.8 & 8.1 & 12.6 & 3.3 & 1.54 & 430.0\\
\hline
\end{tabular}
\end{center}
\end{table}

At the end of the dynamical simulation (t$\sim$
240-260 fm/c) the fragments are well separated, and still bear an average
excitation energy of $\sim$3 \mpn{}, as seen in Table~\ref{tab_BOB}. 
Note that this value is, as expected, smaller than that quoted in 
section~\ref{SIMON} where the simplified assumption that
``gas particles'' were bound in the hot fragments was made.
The mass of the fragments accounts for about 80\% of the total system mass;
at this stage, the experimental observations reported in section~\ref{XeGd}
are practically fulfilled: the average charge of fragments for the two systems is
the same ($<Z>\thicksim13$) and the average multiplicities are in the ratio
of 1.6, which corresponds to the ratio of total masses bound in fragments.
The average radial energy reported in Table~\ref{tab_BOB} for the 
\gu{} system is larger than the extra collective energy introduced in 
the SIMON event generator in section~\ref{SIMON} in
order to account for the average kinetic energies of the fragments. 
The Coulomb energy is however slightly smaller; this stresses the 
difficulty to experimentally determine small expansion energies, 
when their magnitude is similar to the amount of Coulomb energy at the
freeze-out.

\begin{figure}[htbp!] 
\begin{center}
\includegraphics[width=.75\textwidth]{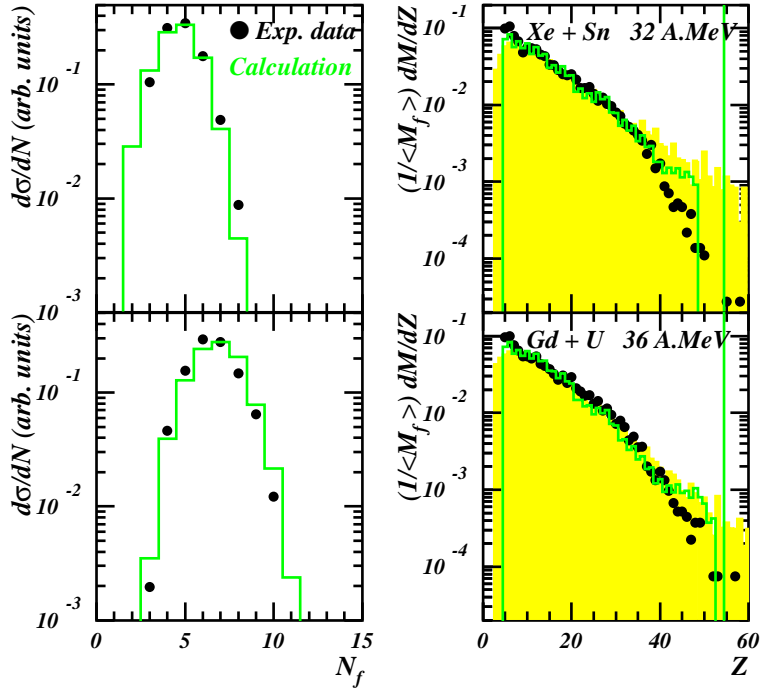}
\caption{\textit{Experimental (points) and simulated (lines) fragment
(Z$\geq$5) multiplicity and charge distributions for single 
multifragmenting sources formed in 32 \mpn{} \xe\ (top) and 36  \mpn{} \gu{}
(bottom) collisions~\cite{FR98}. The filled grey histograms refer to the hot primary
fragments (see Table~\ref{tab_BOB}). \label{BOB1}}}
\end{center}\end{figure}
\begin{figure}[htbp!]
\begin{center}
\includegraphics[width=.75\textwidth]{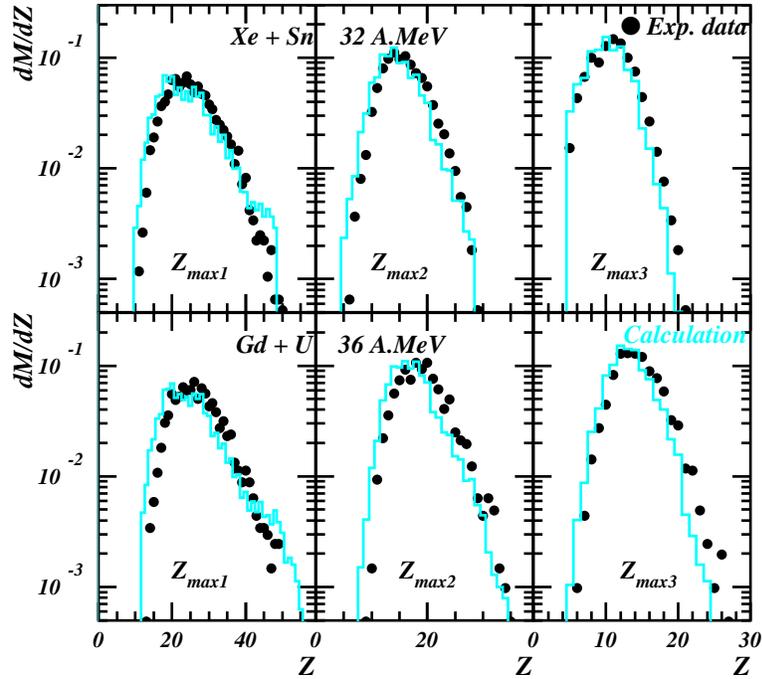}
\caption{\textit{Experimental (points) and simulated (lines) charge
distributions of the 3 largest fragments for single multifragmenting
sources formed in 32 \mpn{} \xe\ and 36  \mpn{} \gu{}
collisions.  \label{BOB2}}}
\end{center}\end{figure}
\begin{figure}[htbp!]
\begin{center}
\includegraphics[width=.75\textwidth]{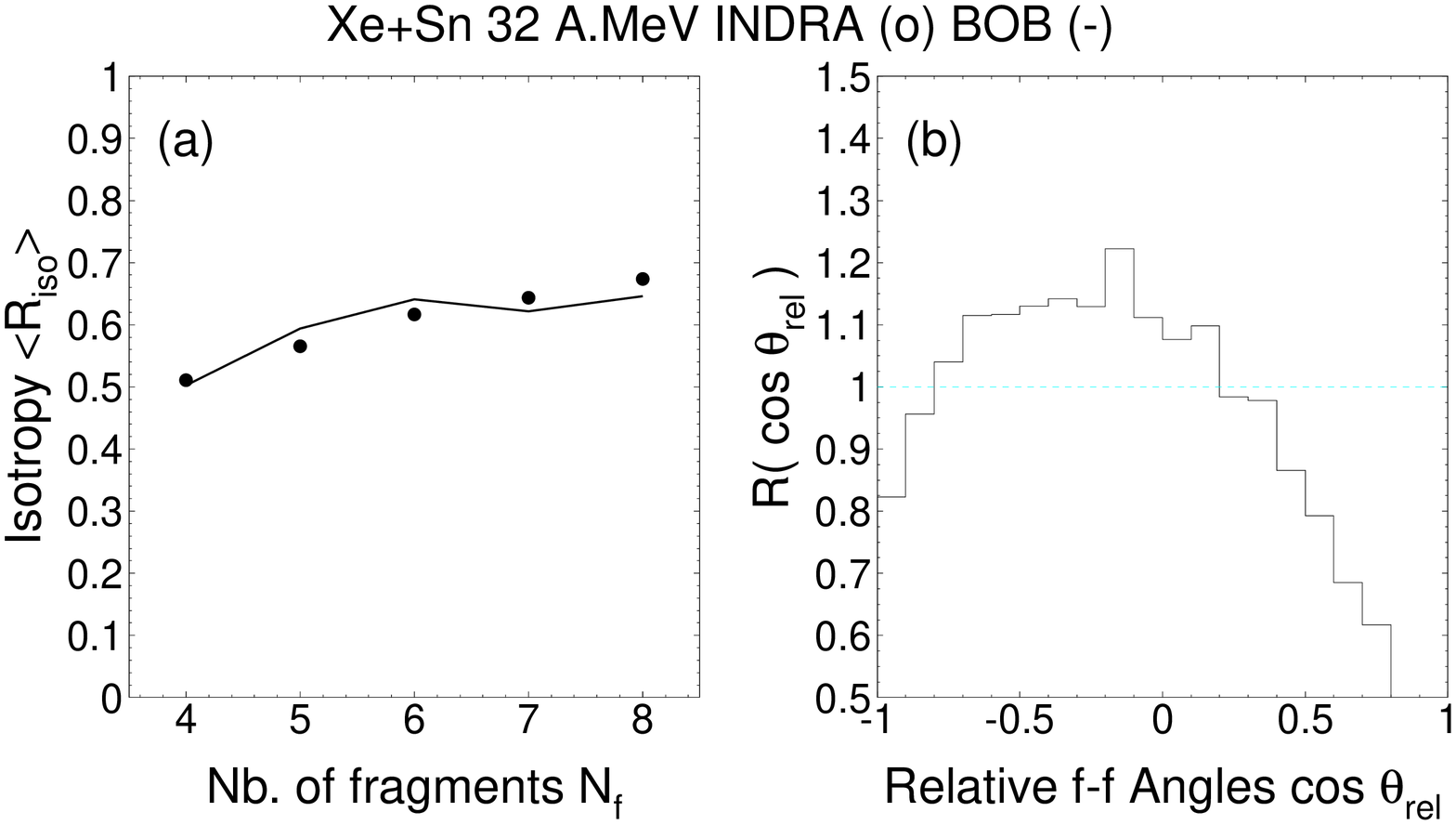}
\includegraphics[width=.75\textwidth]{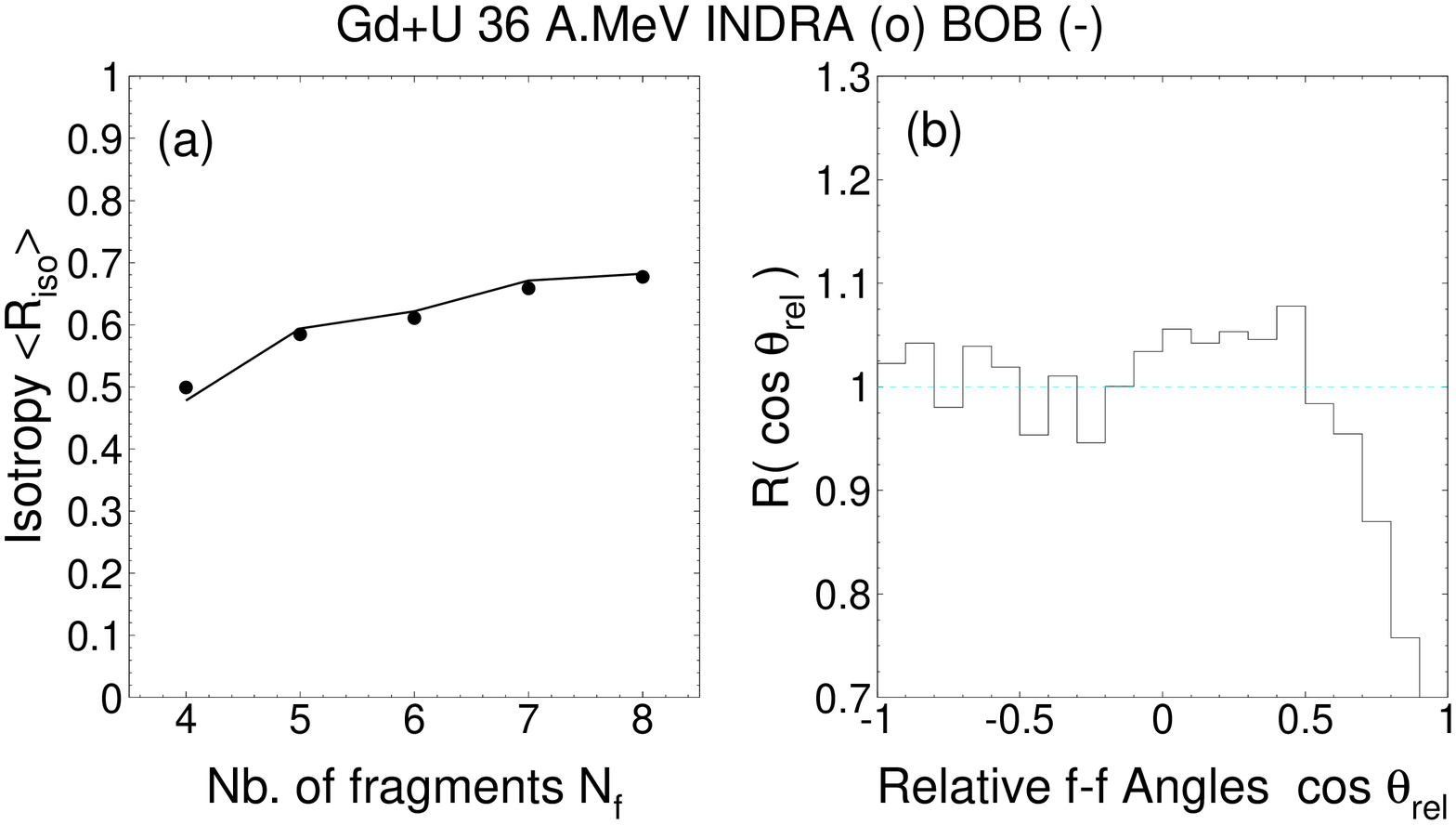}
\caption{\textit{(a)Mean isotropy ratio versus the fragment multiplicity for
experimental (points) and simulated (lines) events. (b) Ratio of the
calculated distributions of the relative angles between fragments to the
experimental ones. A constant ratio at 1 (dashed line) would be obtained
for equal distributions~\cite{FR98}. \label{BOB_forme}}}
\end{center}\end{figure}
\begin{figure}[htbp!]
\begin{center}
\includegraphics[width=.75\textwidth]{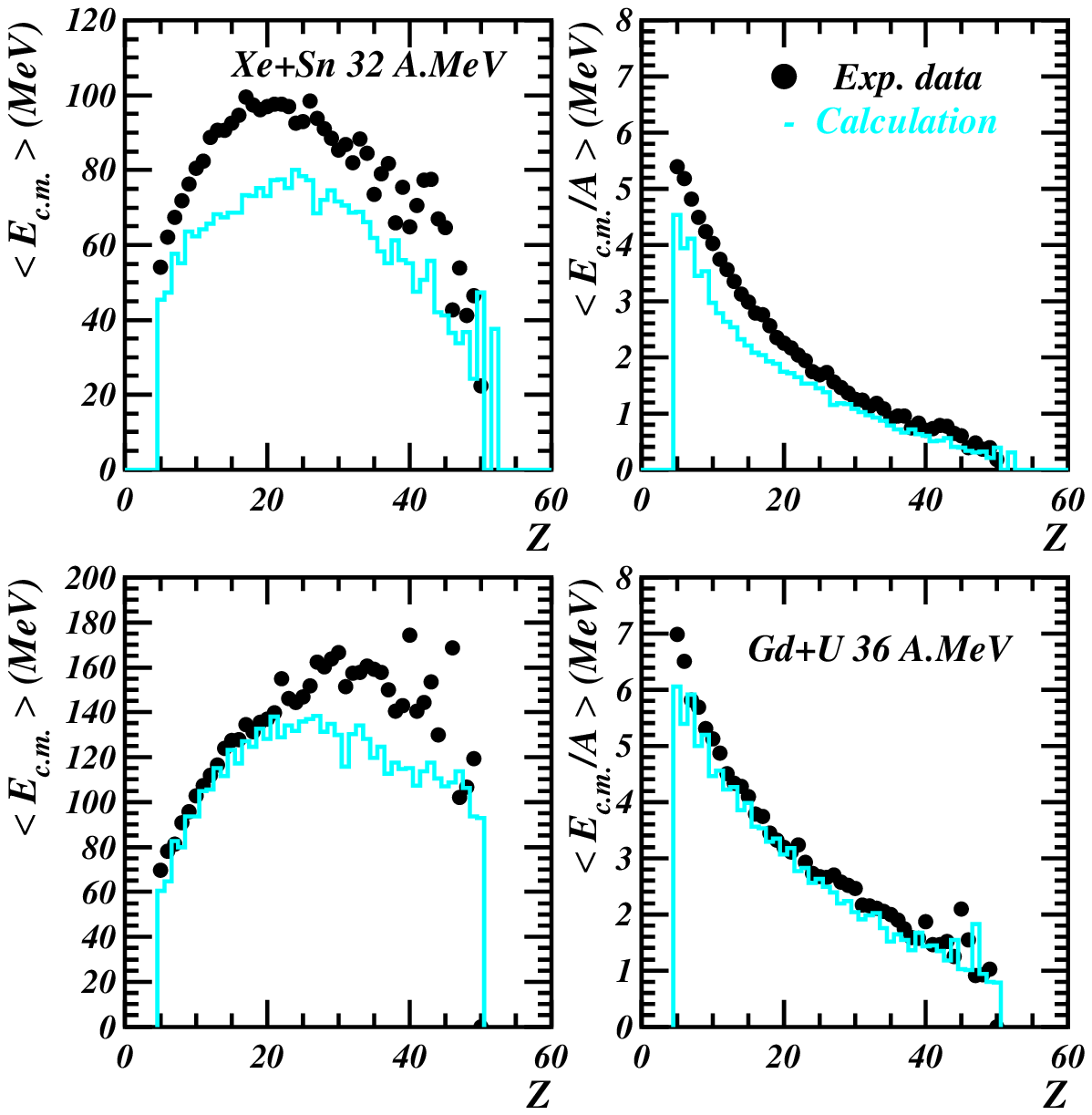}
\caption{\textit{Experimental (points) and simulated (histograms)
average fragment kinetic energies.  The calculated values take into account 
the thermal fluctuations (see text). The results are for
total energies on the left, and in A.MeV on the right. \label{BOB3}}}
\end{center}\end{figure}

In a second step of the simulation, the spatial configuration of the
primary fragments, with all their characteristics as given by BOB, 
is taken as input configuration in the SIMON code (see sect.~\ref{SIMON}): 
the de-excitation of the hot primary fragments is
thus followed while preserving the space-time correlations of all
emitted products. The last step consists in filtering the events to
take into account the experimental set-up and then selecting events with
large flow angles as for the data. It must be noted that the algorithm for
fragment reconstruction excludes light products, and thus all light charged
particles emitted before the freeze-out are lost. This has two consequences:
firstly the selection of complete events is not made on the total detected
charge, but on $Z_{bound}$ (see section \ref{SSE}); experimental data on
Figs.~\ref{BOB1}-\ref{BOB3} use the same selection. Comparisons with the
corresponding figures from section~\ref{SSE} show that the two
selections give identical results. Secondly, no comparison can be made
between calculated and experimental light charged particle properties.
Figure~\ref{BOB1} shows that the calculated multiplicity and charge
distributions of fragments well match the experimental data. The role of
secondary decay in the fragment distribution is negligible, as observed
by comparing the grey  (primary fragments) and the solid (final
fragments) histograms. More detailed comparisons of the charge
distributions of the three largest fragments display a good
agreement (Fig.~\ref{BOB2}): the increasing difference between the
average charges for the two systems when going from the largest to the
second and the third largest fragments is accounted for by the
simulation.
The shapes of the events expressed for instance through the isotropy
ratio, with its multiplicity dependence, are identical in the experiment
and in the simulation, as shown in Fig.~\ref{BOB_forme}a. Conversely the
angular correlations between fragments presented in Fig.~\ref{BOB_forme}b
are less populated at small relative angles in the calculation. This
could suggest that the Coulomb effect is too strong in the calculation
(longer real emission times?). The same effect would also be obtained
if the freeze-out shape resembled a bubble, but in this case
the event shape and the evolution of the fragment energies
on their charge would be different. The experimental selection can also
retain some deformed events~\cite{BOU00}, which are not accounted for in the
simulation where spherical symmetry was kept. The disagreement between 
experiment and calculation on this point  is not fully explained.
Finally the most crucial test is on the fragment kinetic energies,
displayed in Fig.~\ref{BOB3}. Let us recall that as compared to
ref.~\cite{RI98}, the energy calibration of the INDRA CsI located past
45$^o$ (lab) were improved for fragments with charge larger than
15 (see Fig.~\ref{ek_on}). This correction is
particularly important when the velocity of the reaction centre of mass
is small, and the fragment velocities in the c.m. are large. Namely the
new results are very different for the 36 \mpn{} Gd+U
and practically unchanged for the 32 \mpn{}
Xe+Sn central collisions. The improved simulations now predict the
fragment energies for the Gd+U system with  a good  agreement;
for Xe+Sn, the calculated energies fall $\sim$ 20\% below the measured 
values; this remains satisfactory if one remembers that there were 
no adjustable parameters in the simulation. 

\section{Conclusions}\label{conc}

The properties of the single multifragmenting source formed in central 36
\mpn{} \gu{} collisions were studied in detail. The angular and energy
distributions of all fragments and of most of the light charged particles 
are isotropic in the c.m. This indicates that the source has reached
thermal equilibrium when it multifragments. Fragment kinetic energies sign
the onset of expansion energy around 30 \mpn{}.

The measured charge distribution is identical to that obtained for the
lighter \xe{} system with the same available energy, while the average
fragment multiplicities are in the ratio of the total charges of the
systems. This independence of the Z distribution, experimentally observed
for the first time, can be considered as strong evidence of a bulk effect
in the production of these fragments. Note that this observation naturally 
breaks down when the total charge of the systems gets smaller than 
$\sim$60, which is the maximum measured fragment charge for the heavy
systems considered here.

This experimental observation can be related either to bulk instabilities in
the liquid-gas coexistence of nuclear matter (spinodal instabilities) or
taken as a signature of a large exploration of phase space for such heavy
systems. Indeed multiplicities, charge distributions and average fragment 
kinetic energies are equally well predicted by dynamical (BNV-BOB) and
statistical (SMM) simulations~\cite{SA97,NLN99,BOU99}.

Within the dynamical approach, a complete scenario of these central 
collisions may then be proposed: a gently compressed system expands 
and reaches thermal equilibrium at about the time it enters the 
spinodal region. There, the development of density fluctuations causes 
the disassembly of the system into many (hot) fragments and particles. 
When the fragments get free from the nuclear force, they are in a 
``freeze-out'' configuration, and only subject to Coulomb repulsion.
At that time, $\sim $250 fm/c in the simulations, the system has explored 
enough of the phase-space in order to be describable through statistical 
models.  Within such a scenario, there is no contradiction between a
``dynamical'' and a ``statistical'' approach; the first one completely 
describes the time evolution of the collision, and thus helps in 
learning about nuclear matter 
and its phase diagram. The second starts from the phase diagram and has 
more to do with the thermodynamical description of finite nuclear systems. 

Ultimate constraints on models can be expected from the study of 
correlations: fragment velocity and size correlations will allow to 
trace back to the fragment topology at freeze-out, and to look for 
possible fingerprints of spinodal decomposition. This work is in progress.



\end{document}